\documentstyle[prb,aps,multicol,graphics,amsmath,amssymb]{revtex}

\begin{document}
\title{On the two-dimensional metallic state in silicon-on-insulator structures}
\author{G.~Brunthaler$^{a}$\footnote{Corresponding author: Tel.:
+43-732-24689603; Fax.: +43-732-2468-8650; e-mail:
G.Brunthaler@hlphys.uni-linz.ac.at}, A.~Prinz$^{a}$,
G.~Pillwein$^{a}$,
P.E.~Lindelof$^{b}$, J.~Ahopelto$^{c}$}
\address{$^{a}$ Institut f\"{u}r Halbleiterphysik, Johannes Kepler
Universit\"{a}t, A-4040 Linz, Austria\\
$^{b}$ Niels Bohr Institute,
University of Copenhagen, DK-2100 Copenhagen, Denmark\\
$^{c}$ VTT Centre for Microelectronics, Tekniikantie 17, FIN-02150
Espoo, Finland}
\date{\today}
\maketitle

\begin{abstract}
It is shown that the electronic conduction in silicon-on-insulator
(SOI) layers exhibits a metallic regime which is very similar to
that in high-mobility Si-metal oxide semiconductor structures
(MOS). The peak in the electron mobility versus density, the
strong drop in resistivity and the critical concentration for the
metal-insulator transition are all consistent. On the basis of our
SOI data for the temperature and in-plane magnetic field
dependence of the resistivity, we discuss several models for the
metallic state in two dimensions. We find that the observed
behavior can be well described by the theory on the interaction
corrections in the ballistic regime. For the investigated regime,
the temperature dependent screening of scattering potentials gives
also a good description of the data.
\end{abstract}


\begin{multicols}{2}

\section{Introduction}

The discovery of an apparent metallic state in Si-metal oxide
semiconductor (MOS) structures \cite{Krav94+95} has attracted much
attention as it seemed to contradict the two-dimensional (2D)
localization behavior \cite{Abra79}.  It was found that above a
critical carrier concentration $n_c$, the resistivity strongly
decreases with decreasing temperature $T$ (metallic regime),
whereas it increases for lower densities $n < n_c$ (insulating
regime).  The scaling parameter $T_0$ shows a critical behavior
around $n_c$ \cite{Krav94+95}.

At first the discovery of the metallic state in two dimensions was
met with great scepticism.  But the metal-to-insulator transition
(MIT) was confirmed first in Si-MOS structures from a different
source \cite{PopovicPRL97} and further in several other 2D
material systems, which are n- and p- SiGe and AlGaAs and n-AlAs
structures (for references see \cite{Abrahams2000RMP}).

After these experimental findings, the question was raised whether
the metallic behavior manifests a new electronic ground state with
dominating quantum behavior or if the strong resistivity drop
towards low $T$ is based on familiar (i.e.\ non coherent) effects
(for a detailed discussion see \cite{Abrahams2000RMP} and
references therein).

In this work, we report on the finding of a metallic state in
silicon-on-insulator (SOI) structures. In
section~\ref{sec:Experiment}, on the experiment, we describe the
investigated samples and the setup. We further present the data on
the temperature and the magnetic field dependence of the
resistivity and we show the mobility versus density behavior. The
experimental results of the SOI system are compared in detail with
the properties of Si-MOS structures and the differences are
discussed. The discussion of the results follows in
section~\ref{sec:Discussion}. Here we discuss and compare five
important suggestions on the physical origin of the metallic state
in two dimensions. It turns out that the coherent interaction
corrections to the conductivity in the ballistic regime
\cite{Zala01a,Zala01b} are able to describe the observed
resistivity behavior quite well. For the investigated regime also
the temperature dependent screening of scattering potentials
\cite{GoldPRB86} gives a fairly good description of the data.
After a short summary, we describe in
appendix~\ref{sec:AppendixTrapModel} our numerical calculations on
the charged trap state model in detail.

\section{Experiment} \label{sec:Experiment}

\subsection{Samples and setup}

\begin{figure} \vspace{0.1cm} \begin{center}
\resizebox{0.85\linewidth}{!}{\includegraphics{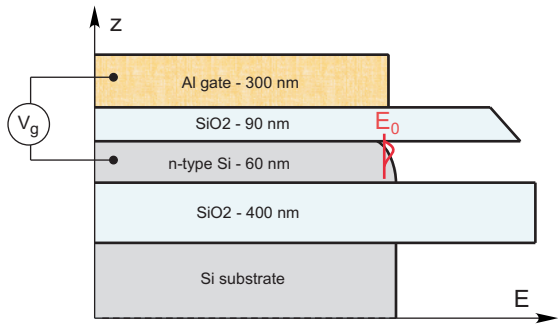}}
\begin{minipage}{8.5cm}
\vspace{0.2cm} \caption{Scheme of the sample structure for
silicon-on-insulator structures. The vertical axis shows the layer
sequence (not to scale), the horizontal axis the band offsets.
\label{fig:SampleScheme} }
\end{minipage} \end{center} \end{figure}

Our investigations were performed on two high-mobility SOI
structures which were prepared recently in Finland. The SOI Hall
bars have identical peak mobilities of $\mu_p =
11,600$\,cm$^2$/Vs.  They were fabricated on commercial 'Unibond'
wafers produced by the Smart-Cut process \cite{Bruel98}.  The
thickness of the Si-layer amounts 200\,nm and that of the buried
oxide layer 400\,nm in these wafers. The background doping is
n-type of about 10$^{15}$\,cm$^{-3}$ which freezes out at low
temperature. After a two-step thermal oxidation, the final
thickness of the Si-layer is 60\,nm.  The gate oxide thickness is
about 90\,nm with a dielectric strength of approximately 1\,V/nm.
The electric contacts were phosphorus-implanted to an activated
carrier concentration of $4\times10^{19}$\,cm$^{-3}$.  The
Al-metallization is 300\,nm thick and sintered at 400$^\circ$C.  A
scheme of the final layer sequence is shown in Figure
\ref{fig:SampleScheme}.  The etched Hall bars have a size of
800\,$\mu$m times 100\,$\mu$m giving a length to width ratio of 8.
All resistivity and Hall measurements were performed in a four
terminal AC-technique at low frequencies of typically 13 or
17\,Hz.  The SOI structures were investigated in a $^4$He cryostat
down to 1.4\,K.  Most investigations were performed on sample
number L2, sample L1 showed in several measurements practically
the same behavior.

The SOI sample number L2 was intensively investigated in a density
range from $1.3\times 10^{11}$ to $1.04\times 10^{12}$\,cm$^{-2}$.
At the low density side of this range, the contacts became high
ohmic. Due to capacitive effects, the measured voltages showed a
delay time of more than 100\,ms before they came into a steady
state. Thus voltage measurements in DC-technique lead to other
values than those with the 13\,Hz AC-technique. We limited our
measurements to the range were the AC- and DC-technique gave the
same results and performed all investigations in the AC-technique
as this gives a better signal-to-noise ratio. At the higher limit
of the investigated density range, the gate started to leak and
the current between gate and sample would have lead to errors in
the resistivity calculations. The investigations were thus kept
in the above given density range.

\vspace{-0.3cm}

\subsection{Resistivity behavior}

From the resistivity and Hall measurements, the carrier density
$n$ and the mobility $\mu$ were calculated assuming the linear
Drude behavior to be valid.  This is not clear a priori as quantum
interference effects might give relevant contributions to the
conductivity and would alter the necessary evaluation.  The
possible importance of quantum effects will be discussed later in
this work.  For now, $n$ and $\mu$ should be used as apparent
Drude parameters for the 2D electron gas.

\begin{figure} \vspace{0.1cm} \begin{center}
\resizebox{0.85\linewidth}{!}{\includegraphics{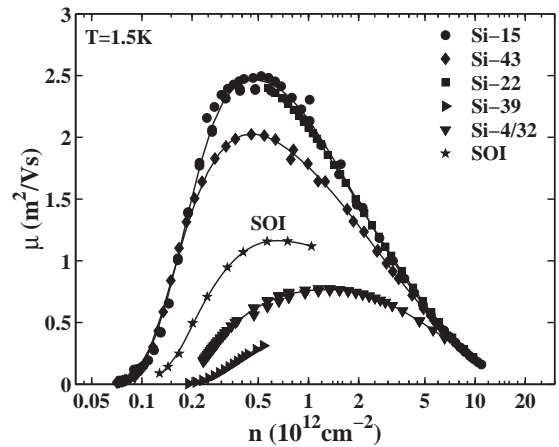}}
\begin{minipage}{8.5cm}
\vspace{0.2cm}
\caption{Mobility $\mu$ versus density $n$ for a
SOI structure in comparison with several Si-MOS samples at a
temperature of 1.5\,K.
\label{fig:mSiSOI} }
\end{minipage} \end{center} \end{figure}

We have evaluated the dependence of mobility $\mu$ on the electron
density $n$ by varying the gate voltage $V_g$ at a constant
temperature of $T = 1.5$\,K. In Fig.~\ref{fig:mSiSOI}, the
$\mu(n)$-dependence for SOI is compared with that for several
Si-MOS samples \cite{Pudalov01Condmat} with different mobilities.
The SOI structure has a peak mobility $\mu_p$ of
$11,600$\,cm$^2$/Vs at a density of $6.5\times
10^{11}$\,cm$^{-2}$. As can be seen from Fig.\ \ref{fig:mSiSOI},
the Si-MOS samples have lower or higher peak mobilities, depending
on the sample quality, but the overall behavior of $\mu(n)$ is
very similar for the two sample types.

At high densities, the electron wave function is squeezed by the
strong electric field of the triangular potential towards the
gate-sided Si/SiO$_2$ interface and thus interface roughness
scattering dominates. At low $n$, the dominating scattering
process is caused by impurities. At the transition region between
the two scattering mechanisms, the mobility reaches its peak value
$\mu_p$ \cite{GoldPRL85}.

\vspace{-0.3cm}

\begin{figure}
\begin{center}
\resizebox{0.95\linewidth}{!}{\includegraphics{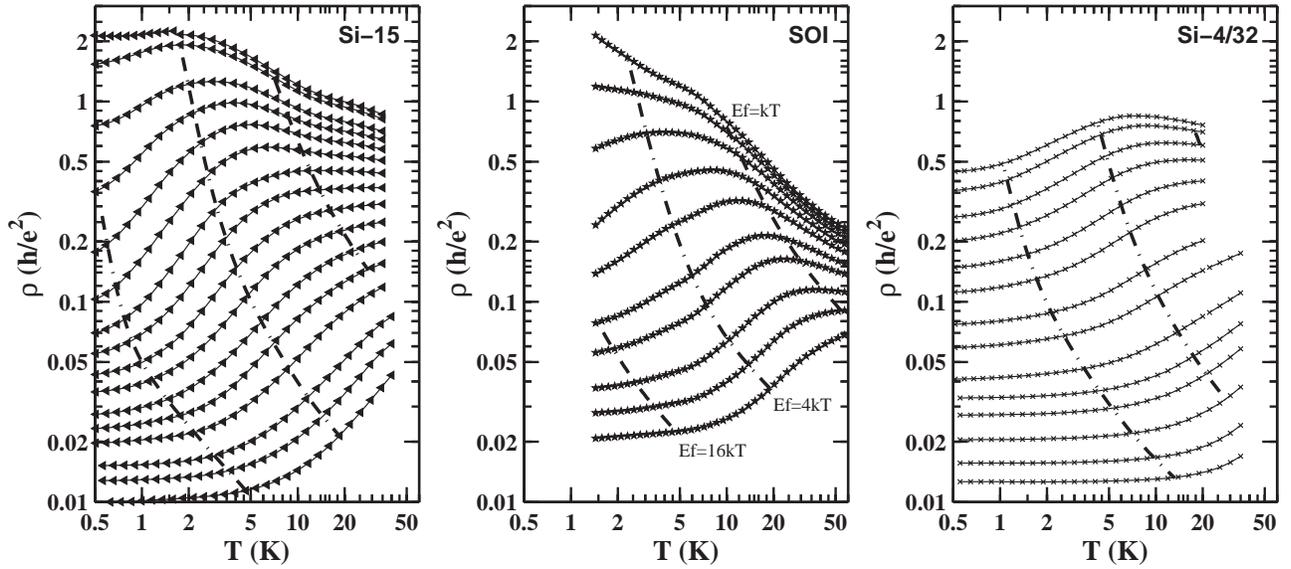}}
\begin{minipage}{16cm}
\vspace{0.2cm} \caption{Resistivity versus temperature behavior in
units of $h/e^2$ for a) high-mobility Si-MOS sample Si-15 ($\mu_p
= 2.7$\,m$^2$/Vs), b) SOI-L2 sample (1.17\,m$^2$/Vs), and c)
medium mobility Si-MOS Si-4/32 (0.8\,m$^2$/Vs). The electron
densities for Si-15 are between 0.93 and $10.2\times
10^{11}$\,cm$^{-2}$, for SOI-L2 are 1.28, 1.44, 1.67, 2.02, 2.47,
3.26, 4.06, 5.65, 7.50 and $10.4\times 10^{11}$\,cm$^{-2}$ and for
Si-4/32 between 2.34 and $28.9 \times 10^{11}$\,cm$^{-2}$.
\label{fig:rSiSOI} }
\end{minipage}
\end{center}
\end{figure}

The most typical feature of the metallic state in 2D systems is
expressed by the strong resistivity drop towards lower
temperatures.  This strong drop in $\rho(T)$ is clearly visible
in Fig.\ \ref{fig:rSiSOI}b for the SOI sample in the range of
$1.67\times 10^{11} \leq n \leq 1.04\times 10^{12}$\,cm$^{-2}$.
The maximum decrease in $\rho(T)$ amounts up to about a factor
3.5 for $n = 2.47\times 10^{11}$\,cm$^{-2}$.  For the two lower
density curves with $n = 1.28$ and $1.44\times
10^{11}$\,cm$^{-2}$, an insulating behavior is observed.  The
critical concentration $n_c$, at which the behavior changes from
insulating to metallic, lies at about $1.55\times
10^{11}$\,cm$^{-2}$.

For comparison, in Fig.\ \ref{fig:rSiSOI}a and \ref{fig:rSiSOI}c
the very high and medium mobility Si-MOS samples \cite{Bru01PRL}
Si-15 and Si-4/32 with $\mu_p = 27,000$ and $8,000$\,cm$^2$/Vs,
respectively, are shown. In sample Si-15, the strong decrease in
$\rho(T)$ is shifted to lower $T$ and the maximum decrease amounts
up to a factor 7, in Si-4/32 the decrease is at somewhat higher
$T$ in comparison to the SOI structure and the maximum decrease is
about a factor 3.  This shows a clear trend of the properties of
the metallic regime with the sample quality, i.e.~the peak
mobility $\mu_p$ and manifests the similarities between SOI and
Si-MOS structures. Additionally, in Fig.\ \ref{fig:rSiSOI},
dashed-dotted lines at $E_f = k_BT$, $4k_BT$ and $16k_BT$ are
shown.
\end{multicols}

\vspace{-0.3cm}

\begin{figure}
\vspace{0.1cm}
\begin{center}
\resizebox{0.90\linewidth}{!}{\includegraphics{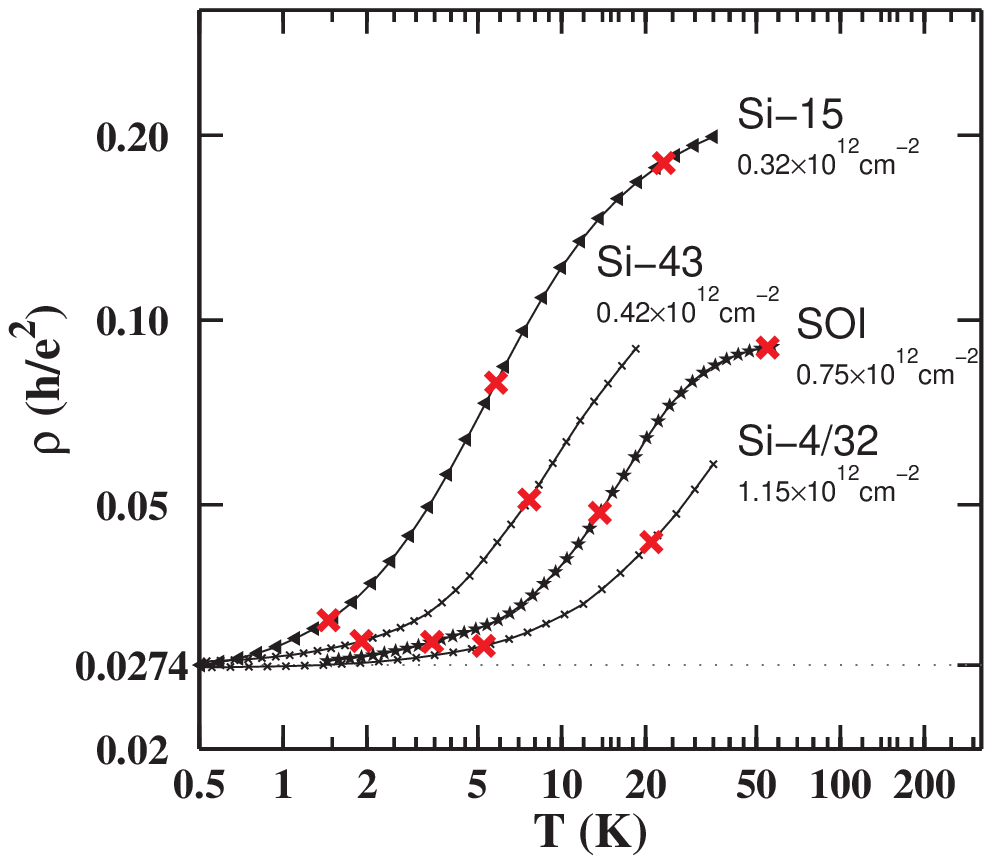}}
\begin{minipage}{8.5cm}
\vspace{0.2cm} \caption{$\rho(T)$ for samples Si-15, Si-43, SOI-L2
and Si-4/32 for one and the same low-$T$ resistivity of $0.0274
h/e^2$. The crosses on the curve from left to right mark the
temperature where $k_BT = E_F/16$, $E_F/4$, and $E_F$.
\label{fig:rSiSOIsingle} }
\end{minipage}
\end{center}
\end{figure}

\begin{multicols}{2}
Figure \ref{fig:rSiSOIsingle} shows a direct comparison of the
strong changes in $\rho(T)$ for the four samples Si-15, Si-43,
SOI-L2 and Si-4/32 for the same low-$T$ resistivity $\rho = 0.0274
h/e^2$. The same behavior, as described above, is visible: the
higher the peak mobility, the stronger is the change in $\rho(T)$
and the lower is the temperature where the increase starts.

\vspace{-0.3cm}

\begin{figure}
\vspace{0.1cm}
\begin{center}
\resizebox{0.80\linewidth}{!}{\includegraphics{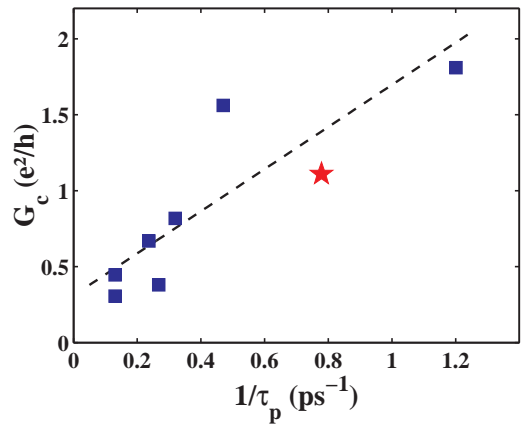}}
\begin{minipage}{8.5cm}
\vspace{0.2cm} \caption{Critical conductance $g_c$ in units of
$e^2/h$ versus the inverse peak momentum relaxation time
$1/\tau_p$ for the SOI structure L2 (star symbol) and several
Si-MOS samples (square symbols). The dashed line shows a linear
regression curve. \label{fig:gctp} }
\end{minipage}
\end{center}
\end{figure}

For Si-MOS structures, it was shown that the critical conductance
$g_c = 1/\rho_c$ changes systematically with the peak momentum
relaxation time $\tau_p = m^* \mu_p/e$, which is a measure of the
sample quality \cite{Pudalov98PhysE}. Here $m^*$ denotes the
effective conductivity mass which is the transverse mass of the
two lower lying valleys of $m^*_t = 0.19$\,$m_e$ and $e$ the
elementary charge.
For the SOI structure, $g_c$ is about $1.1 e^2/h$ and is compared
in Fig.\ \ref{fig:gctp} with the values for several Si-MOS samples
\cite{Pudalov98PhysE}. It can be seen, that the $g_c$-value for
SOI coincides quite well with the general behavior of the Si-MOS
samples.

\vspace{-0.3cm}

As was seen so far, the SOI structure behaves very similar to a
high mobility Si-MOS sample.  Therefore, we ask the question: What
are the similarities and differences between the two sample types?
Due to the different effective masses $m^*$ perpendicular to the
Si/SiO$_2$ interfaces and due to strain effects, the six-fold
conduction band (CB) degeneracy at the $\Delta$-points of bulk Si
is lifted and only the two lower lying valleys with their
longitudinal axis perpendicular to the 2D layer are occupied. The
Si layer in SOI is only 60\,nm thick, below there is the buried
oxide layer, which constitutes also of a thermally grown oxide
fabricated before the wafer bonding process. The active Si-layer
has an n-type background doping of about $10^{15}$\,cm$^{-3}$
which freezes out at low $T$.  In the Si-MOS samples, the
electrons are confined in the surface inversion layer of a
nominally p-type doped bulk Si crystal.  Our high mobility Si-MOS
samples were doped with an acceptor density of about $2\times
10^{15}$\,cm$^{-3}$.

Due to the applied gate voltage $V_g$, an approximately triangular
potential well results for the electrons at the upper Si/SiO$_2$
interface.  Nevertheless, the extension of the electronic wave
function is different for the SOI and the Si-MOS structure due to
the different background doping.  In the n-type SOI structure, the
Fermi level $E_F$ is pinned underneath the 2D layer between the
donor level and the CB edge. The band bending below the 2D layer
is therefore quite small. In p-type Si-MOS, there is a wide
depletion layer of about 0.85\,$\mu$m below the 2D layer, as $E_F$
is pinned here between the acceptor level and the valence band
(VB) edge in the undistorted p-type region. The different band
bending below the 2D layer has an influence on the extension of
the wave functions especially for small electron densities.

\begin{figure}
\vspace{0.1cm}
\begin{center}
\resizebox{0.85\linewidth}{!}{\includegraphics{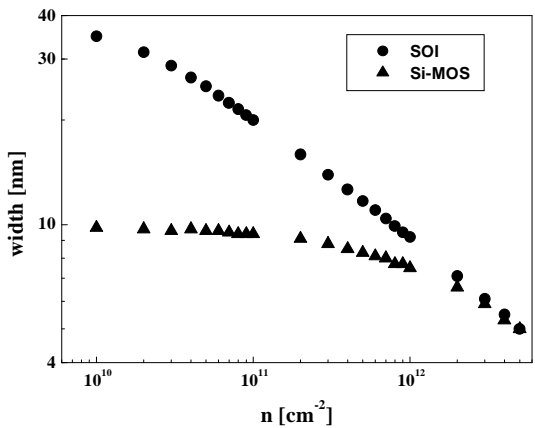}}
\begin{minipage}{8.5cm}
\vspace{0.2cm} \caption{Width of electron wavefunction versus
electron density $n$ for SOI (circle symbol) and Si-MOS
(triangle). The width is defined to include the center 90 percent
expectation probability of the wave function, 5 percent at the
left and the right hand side are omitted. \label{fig:PsiWidth} }
\end{minipage}
\end{center}
\end{figure}
\vspace{-0.3cm}

We have calculated the wave function $\Psi(z)$ and the triangular
potential $V(z)$ by solving the Schr\"{o}dinger and Poisson
equations in a self consistent way (with $z$ the coordinate
perpendicular to the interface layer).  In order to solve the
Schr\"{o}dinger equation, the charge distribution $\rho(z) \propto
|\Psi^2(z)|$ was taken into account.  No exchange or correlation
effects among the electrons were considered, thus giving a first
order approximation of $\Psi(z)$ and $V(z)$. Figure
\ref{fig:PsiWidth} shows the width of the wave function versus
electron density $n$.  The width of $\Psi$ is defined as the
length where the central 90 percent of the expectation probability
$|\Psi^2|$ is contained, with 5 percent at the left and at the
right side ignored. For a high electron density of
$10^{12}$\,cm$^{-2}$ the width is with 7.6 and 9.2\,nm nearly the
same for both structures.

At lower densities, the difference in the width of $\Psi$
increases.  The depletion width of 0.85\,$\mu$m at an acceptor
density of $2\times 10^{15}$\,cm$^{-3}$ corresponds to a 2D
charge density of $1.7 \times 10^{11}$\,cm$^{-2}$.  Thus, at that
2D electron density, the electric field at the Si/SiO$_2$
interface is in Si-MOS twice as high as in the SOI structure.  As
can be seen from Fig.\ \ref{fig:PsiWidth}, the difference in the
wave function extension is indeed large at that density.  At
still lower density, the width saturates in Si-MOS, as the field
from the conducting 2D electrons is less important than that from
the acceptors in the depletion layer.  On the opposite, the width
in the SOI structure further increases, as there is no electric
field confinement from other charges than the conducting
electrons itself.

\begin{figure}
\vspace{0.1cm}
\begin{center}
\resizebox{0.80\linewidth}{!}{\includegraphics{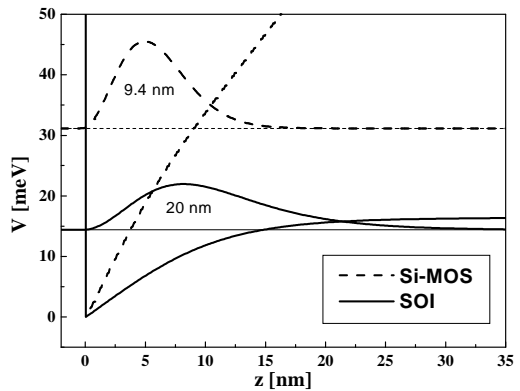}}
\begin{minipage}{8.5cm}
\vspace{0.2cm}
\caption{Comparison of the wave functions $\Psi(z)$
and the potential shape $V(z)$ for the SOI (solid line) and Si-MOS
structure (dashed line) at an electron density of $1\times
10^{11}$cm$^{-2}$. \label{fig:PsiPot} }
\end{minipage}
\end{center}
\end{figure}

Figure \ref{fig:PsiPot} shows for comparison the potential shape
$V(z)$ and the wave function $\Psi(z)$ at $n =
10^{11}$\,cm$^{-2}$.  At that density, the confining potential is
much steeper in Si-MOS and thus the energy $E_0$ of the lowest
electronic subband is with 31\,meV much higher than in SOI with
14\,meV. In Fig.\ \ref{fig:PsiPot}, the difference in the width of
the wave functions is also clearly visible and is caused by the
difference in the potential steepness.  For the case that higher
subbands $\Psi_n(z)$ are occupied, the differences in their
energies and extensions should be even larger.

\vspace{-0.3cm}

In the considered density range, the width of the wave function is
always smaller than the Si-layer thickness of the SOI structure
and thus the electron do not feel the lower Si/SiO$_2$ interface
towards the buried oxide layer.  In order that such an effect
could become significant, the Si layer would have to be as thin as
20\,nm for $n \approx 1 \times 10^{11}$\,cm$^{-2}$ and even
thinner for higher densities.

\subsection{Magnetoresistivity}

\begin{figure}
\vspace{0.1cm}
\begin{center}
\resizebox{0.80\linewidth}{!}{\includegraphics{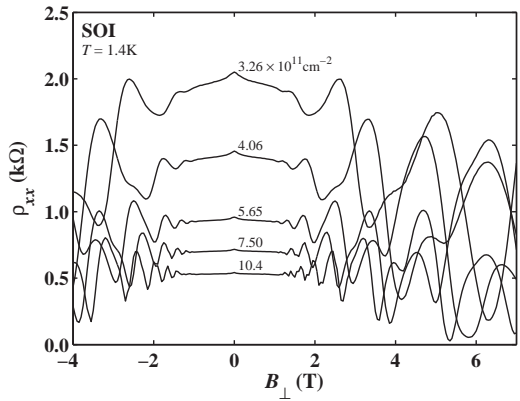}}
\begin{minipage}{8.5cm}
\vspace{0.2cm}
\caption{Magnetoresistivity $\rho_{xx}$ for
magnetic fields oriented perpendicular to the 2D layer for $n =
3.26$, 4.06, 5.65, 7.50, and $10.4 \times 10^{11}$\,cm$^{-2}$ at
a temperature of 1.4\,K. \label{fig:rhoBperp} }
\end{minipage}
\end{center}
\end{figure}

We have also investigated the SOI structures in perpendicular and
parallel (i.e.\ in-plane) magnetic field $B$.  Figure
\ref{fig:rhoBperp} shows typical magnetoresistivity curves
$\rho_{xx}(B_\perp)$ for $B$ perpendicular to the 2D layer for
different densities at $T = 1.4$\,K.  For the lower densities, the
weak localization peak around $B = 0$ is clearly visible.  At
about 1\,T, the Shubnikov - de Haas oscillations set in and gain
strongly in amplitude towards higher magnetic fields. At around
4\,T, the spin splitting is visible.  The behavior in
perpendicular magnetic field is again very similar to that of
Si-MOS samples with comparable mobility.

\vspace{-0.3cm}

\begin{figure}
\vspace{0.1cm}
\begin{center}
\resizebox{0.80\linewidth}{!}{\includegraphics{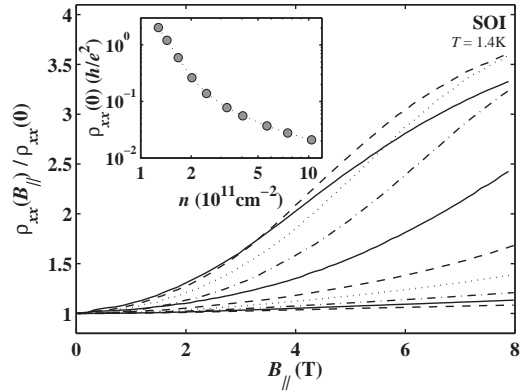}}
\begin{minipage}{8.5cm}
\vspace{0.2cm} \caption{Relative change of the magnetoresistivity
$\rho_{xx}(B_\parallel)/\rho_{xx}(0)$ for magnetic fields
oriented parallel to the 2D layer for
$n = 1.28$, 1.44, 1.67, 2.02, 2.47, 3.26, 4.06, 5.65, 7.50, and $10.4 \times 10^{11}$\,cm$^{-2}$
(from top to bottom) at a temperature of
1.4\,K. \label{fig:rhoBpar} }
\end{minipage}
\end{center}
\end{figure}

For in-plane magnetic field, the relative change
$\rho_{xx}(B_\parallel)/\rho_{xx}(0)$ of the magnetoresistivity of
SOI sample L2 is shown in Fig.~\ref{fig:rhoBpar} for different
densities at $T=1.4$\,K. At high densities of about $1 \times 10^{12}$\,cm$^{-2}$,
the changes are quite small, whereas for the
lower densities there is a strong increase in $\rho_{xx}$ by up to
a factor 4. At the lowest density of $1.28 \times 10^{11}$\,cm$^{-2}$,
the observed change is again lower than for
the previous densities as can be seen from the crossing of the
upper curves in Fig.\ \ref{fig:rhoBpar}. This indicates a
saturation of the strong increase in $\rho_{xx}(B_\parallel)$ at
about $1.5 \times 10^{11}$\,cm$^{-2}$.

An even stronger $\rho(B_\parallel)$-dependence was observed in
high-mobility Si-MOS structures in the metallic regime
\cite{Pudalov97JETPL,Simon97PRL}.  Again the changes were stronger
for the lower densities $n$.  The increase with $B_\parallel$ was
first attributed to the spin polarization of the conducting
electrons \cite{Kawaji99PRL} and later to the increase in disorder
caused by the magnetic field \cite{Pudalov01Condmat}.  A strong
change of $\rho(B_\parallel)$ was also observed in p-type
GaAs/AlGaAs structures. \cite{Yoon00PRL}

\vspace{-0.3cm}

\section{Discussion} \label{sec:Discussion}

Above, we have presented new results on the 2D metallic state in
silicon-on-insulator structures and compared its properties with
that of high-mobility Si-MOS samples.  As the physical origin of
the 2D metallic state is still an open question, we like to
discuss the different possibilities in detail.

There appear to remain five important suggestions for the 2D
metallic state at low temperatures.  These are
(i) a new quantum mechanical ground state for electrons in a
solid,
(ii) scattering of electrons at charged hole traps in the oxide
layer,
(iii) interband scattering between different bands,
(iv) temperature dependent screening of impurity scattering, and
(v) coherent interaction corrections in the ballistic
regime.
If the first suggestion proves to be correct, the 2D MIT would
indeed be the manifestation of some new physics, whereas the next
three suggestions are based on settled physical phenomena, which
may be described by terms like conventional, semiclassical or
non-coherent (on the first sight).  The fifth suggestion was
proposed quite recently and it has to be discussed whether it can
be classified in the terminology used above or not.

\subsection{New quantum mechanical ground state}

The first suggestion (i) of a new quantum mechanical ground state
is based on the
scalability of the resistance in the neighborhood of the apparent
MIT in high-mobility Si-MOS structures \cite{Krav94+95}.  It was
found that the $\rho(T,n)$-curves for $n<n_c$ and for $n>n_c$ can
be brought to coincidence on two distinct curves for the
insulating and the metallic regime, respectively, by scaling the
temperature with $T/T_0(n)$.  The scaling parameter exhibits a
power law $T_0 \propto |\delta_n|^b$, with $\delta_n = (n -
n_c)/n_c$ and $b \approx  1.6$.  Such a critical scaling behavior
indicates the metal-insulator transition to be a quantum phase
transition \cite{Dobro97,Abra96}.

The seminal paper on the absence of quantum diffusion in two
dimensions \cite{Abra79} was based on the assumption of
non-interacting electrons.  In this case the weak localization
contribution decreases the conductivity and leads to continuing
higher resistivities with decreasing temperatures.  In the recent
quantum-phase-transition models for the unexpected MIT in 2D, the
electron-electron (e-e) interaction plays a dominant role
\cite{Finkel84,Castellani84,Abra96,Dobro97}.  The quantum coherent
interaction corrections consist of positive and negative
contributions to the conductivity and may lead to a decrease of
the resistivity under certain conditions\cite{Punnoose01}. In any
case, quantum coherence is a prerequisite that the MIT can be a
quantum phase transition.

A metallic behavior, i.e.\ a strong drop in the resistivity, was
not only observed in the near vicinity of the critical
concentration $n_c$ but also far away, for densities up to $n
> 1\times10^{12}$\,cm$^{-2}$ (i.e.\ more than 10 times $n_c$).
As the decrease in $\rho(T)$ looks quite similar near $n_c$ and
far away, it is quite instructive to
look for quantum effects over an extended density range.  In a
recent work \cite{Bru01PRL}, the temperature dependence of the
weak localization was investigated in the metallic state of
high-mobility Si-MOS structures in order to find the borders of
quantum coherence. In that work, the temperature threshold $T_q$
for the single-electron quantum coherence was deduced from the
crossing point of the temperature dependence of the phase
coherence time $\tau_\varphi$ and the momentum relaxation time
$\tau$. When $\tau_\varphi > \tau$, the coherence time is long
enough that an electron can return in the coherent state to its
origin whereas for $\tau_\varphi < \tau$ this is not possible.

It was found for the Si-MOS system \cite{Bru01PRL} that for $n
> 2\times10^{11}$\,cm$^{-2}$, the strong decrease in
$\rho(T)$ takes place above $T_q$ and thus should not be related
to a single-electron coherent effect.  Also the second border $k_B
T_{\rm ee} = \hbar/\tau$ lies below the strong resistivity drop at
still higher densities.  Therefore also the e-e interaction
induced quantum corrections to conductivity should not be
responsible for the resistivity drop in this regime.  This should
at least be deduced in the picture where quantum coherent effects
and screening are attributed to two different physical phenomena.
But recently an interesting paper by Zala et al.\ has appeared
\cite{Zala01a}, which may indicate that the two before mentioned
effects are caused by the same physics.
This will be discussed in more detail later in this work.

In any case, the work on the borders for quantum coherence
\cite{Bru01PRL} is strictly applicable only in the higher density
regime and does not give a direct answer on the importance of
quantum coherence in the close vicinity of the critical density
$n_c$.
In a recent work, Punnoose and Finkelstein show \cite{Punnoose01}
by a renormalization group analysis that especially for
high-mobility Si-MOS structures, the presence of two degenerate
conduction band valleys, stabilizes and enhances coherent e-e
interaction effects in the vicinity of $n_c$. They are able to
describe the temperature dependent changes in conductance in an
$n$-range between 0.83 and $0.94\times10^{11}$\,cm$^{-2}$ without
any adjustable parameter. So it is quite surprising that at low
$n$ the whole drop in $\rho$ seems to come from quantum coherent
interaction effects while at high $n$ the drop takes place without
quantum coherence and nevertheless the strong $T$-dependence looks
so similar in the different regions. It is not yet clear whether
the metallic and insulating behavior belong to a quantum phase
transition or not. This work focuses on the metallic regime and
will not be able to judge on the metal-to-insulator transition
itself.

\subsection{Charged hole traps}

The second suggestion (ii) on the scattering of electrons at
charged hole traps in the oxide layer by Altshuler and Maslov (AM)
\cite{Altsh99PRL} is prepared especially for Si-MOS structures
with the Si/SiO$_2$ interface on the gated side of the 2D layer.
The same model should be applicable also for our SOI structures,
as an equivalent Si/SiO$_2$ interface is present for the 2D
electron confinement.

It is well known that in the SiO$_2$ interface layer there exist
several types of defects states \cite{Sze81}.  In the
AM model,
it is assumed that a relative large number of hole trap states
exists at a certain trap energy $E_t$, which is spatially
homogeneously distributed in the oxide layer. It is further
assumed that these trap state lie at a well defined energy
$E_{t0}$ if no external electric field is applied. The potential
gradient due to an applied gate voltage $V_g$ causes a linear
increase of the trap energy position.  In order to describe the
total energy of the trap states, also the mirror charge of the
traps inside the 2D electron gas has to be included. This leads to
a down bending or the energetic position towards the interface and
causes a maximum in the total trap energy $E_t(z)$, with $z$ the
position inside the oxide layer (see Fig.~\ref{fig:trap_energy} in
Appendix \ref{sec:AppendixTrapModel}). When the Fermi energy lies
below certain trap states at energies $E_t(z)$ (for low $n$) these
states are positively charged and cause a large scattering
probability for the 2D electrons.  At high densities all trap
states lie below $E_F$ and only due to the Boltzmann occupation
tail, an exponentially small number of traps contributes to the
electron scattering.  The charged trap states form dipoles
together with their mirror charge in the 2D layer which has to be
taken into account in the calculation of the scattering cross
section.

In order to calculate the resistivity, one needs to know how the
Fermi energy $E_F$ varies with the temperature $T$ (in the work
of AM, the chemical potential at $T = 0$ is called the Fermi
energy $E_F$ whereas the chemical potential at $T > 0$ is denoted
by $\mu(T)$; in this work we use $E_F(T)$ instead).  AM suggest
two scenarios for the $T$ dependence of the Fermi energy of the
2D electron gas : (a) it coincides with the Si substrate or (b)
it is decoupled from the substrate.

For case (a), the calculation of $\rho(T)$ by AM show a behavior
which looks quite similar to the observed temperature dependence
near the critical density $n_c$ for high mobility Si-MOS
structures.  The strong increase in $\rho$ with increasing $T$
occurs around $k_B T \approx E_F - E_t$ as then the occupied trap
states are partially emptied (and charged) due to the broadening
of the Fermi occupation function and the scattering rate is
increased accordingly.  For case (a), the insulating and metallic
regions are distributed quite symmetrically around the separatrix
which occurs for $E_F = \max(E_t(z))$.

For case (b), the resistivity is decreasing in both regimes ($E_F
\gtrless E_t$) according to the calculations of AM
\cite{Altsh99PRL} and seems to be unable to explain the
experimentally observed behavior.  As mentioned by AM, the more
realistic case is (b) as the depletion layer isolates the 2D
electron gas from the substrate and both regions should have
different quasi Fermi energies. This is also verified
experimentally \cite{Ando82quasiFermi}.  This situation causes
some difficulty in order to apply the charge trap model to the
high-mobility Si-MOS structures as well as to the SOI samples
considered in this work.

\begin{figure} \begin{center}
\resizebox{0.85\linewidth}{!}{\includegraphics{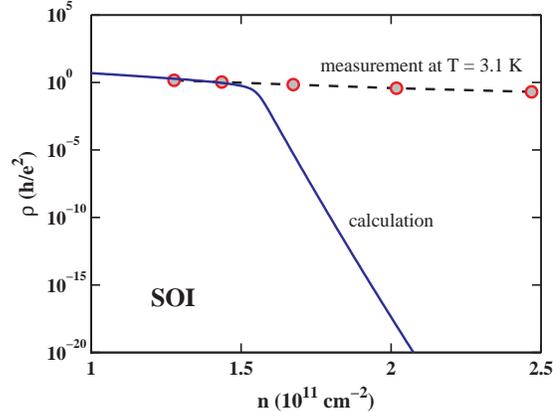}}
\begin{minipage}{8.5cm}
\vspace{0.2cm} \caption{Comparison of $\rho(n)$ at $T_i = 3.1$\,K
for experimental data (circles) and calculations (solid lines)
according to the AM-trap model.
\label{fig:rho_trap_SOI} }
\end{minipage} \end{center} \end{figure}

We have performed numerical calculations of the AM-trap model for
the SOI data.  These calculations are described in appendix A in
detail.  If one applies an increasing gate voltage $V_g$ in order
to increase the electron density $n$ in the 2D layer, the spatial
maximum in the trap energy shifts quite fast to lower energies.
This leads to a strong occupation of the trap states as soon as
the maximum shifts below the Fermi energy $E_F$ and the scattering
efficiency decreases drastically.  In Fig.\
\ref{fig:rho_trap_SOI}, we have calculated the $\rho(T)$
dependence at an intermediate temperature of $T_i = 3.1$\,K, where
we have a set of measurement data.  The temperature $T_i$ was
chosen so that $\rho(n, T_i)$ is for a large density interval
neither in the low-$T$ saturation nor in the ``normal'' high-$T$
regime.  The resistivity behavior should be dominated by thermal
excitation of carriers according to the AM-trap model. The low-$T$
saturation of $\rho(T)$ in the metallic regime cannot be described
with the AM-trap model alone, as all traps get filled according to
the exponential Boltzmann-tail and the resistivity tends to zero.
Only by assuming additional residual scattering centers, the
resistivity saturates.  We have evaluated the measurement data at
$T_i$ only for densities where the low-$T$ saturation should not
have a large influence yet.

Figure \ref{fig:rho_trap_SOI} shows that the decrease in
resistivity vs.\ density is too strong by many orders of magnitude
in the AM-trap model.  At $n = 2.5 \times 10^{11}$\,cm$^{-2}$,
$\rho$ should be below $10^{-30} h/e^2$.  This low value results
from a very small number of positively charged trap states
according to the Fermi-Dirac occupation function. Temperature
dependent screening effects were not considered in this
calculation as they give only a change in $\rho$ by one or two
orders of magnitude, but the calculated resistivity is by about 30
orders of magnitude too small compared with the measured values.
In the same density interval, the measured resistivity decreases
just by about one order of magnitude, which appears as a nearly
flat line on the scale of Figure \ref{fig:rho_trap_SOI}.

\vspace{-0.3cm}

We conclude on the AM-trap model that it cannot explain the
experimental behavior of the SOI (and similarly of the Si-MOS)
samples at intermediate temperatures as the carrier freeze out on
the trap states occurs by far too fast if the trap maximum is
below $E_F$.  Further improvements of the AM-model may be
possible, but one has to note that the addition of trap states at
other energies or the energetical broadening of the trap states
around $E_t$ destroys the critical behavior around $n_c$ and the
modified model would probably not be able to describe the abrupt
metal to insulator transition.

Moreover, case (a) of the trap model by AM seems to be able to
explain the $\rho(T)$ dependence only in a very narrow density
range around $n_c$ but can not explain the metallic behavior for
the wide $n$ range as observed in Si-MOS and SOI samples.  Thus it
seems to be possible that the hole trap model in the present form
may explain some important contributions to $\rho(T)$ in Si-MOS
and SOI but is unable to describe the whole temperature
dependence.  As mentioned above, the AM model is especially
prepared for samples with an SiO$_2$ oxide layer as on border of
the 2D electron gas.
The model will thus probably not be able to explain the
metal-insulator transition in other material systems with strong
changes in $\rho(T)$ like p-type AlGaAs or Si/SiGe. On the other
hand, it is not clear anyway whether the apparent metal-insulator
phase transition has a universal origin or if it is caused by
several different effects in the different material systems.

\subsection{Interband scattering}

The third suggestion (iii) is attributed to interband scattering
between different conduction or valence bands. The correlation
between the existence of two conduction bands and the metallic
behavior was proposed by Pudalov for Si-MOS.\cite{Puda97a} He
assumed that the spin-orbit coupling due to the asymmetry of the
inversion layer potential is strong enough in order to create the
splitting. But the calculation of the spin-orbit coupling is not
straight forward and have to be performed with great
care.\cite{ZawadzkiPRB99+01} A later estimate by Pudalov et
al.~\cite{Pudalov02PRLa} gives a clearly smaller value for the
spin-orbit coupling constant of $\alpha \approx 6 \times
10^{-6}$\,K\,cm, so that the corresponding splitting should not be
the origin of the metallic behavior in Si-MOS structures. For
p-type AlGaAs Papadakis et al.\ have argued
\cite{PapadakisScience99} on the spin-orbit coupling due to the
inversion asymmetry of the electron layer potential, but it turned
out that the effect is due to an anomalous behavior of the
magneto-oscillations \cite{WinklerPRL00}.

Nevertheless, the apparent metallic behavior in p-type AlGaAs
layers
strongly points towards a two-band transport effect.
\cite{Yaish00PRL} There, the heavy hole band is split and
inelastic interband Coulomb scattering between the two bands with
different dispersion takes place.  This was confirmed by
magnetotransport measurements, where the positive
magnetoresistance is attributed to the classical two-band
scattering effect. But for n-type Si-based heterostructures, the
two-band transport should not be of great importance, as the
conduction band splitting seems to be rather small.

\subsection{Temperature dependent screening}

The forth suggestion (iv) on temperature dependent screening
evolves from calculations of the mobility versus electron density
dependence for Si-MOS
\cite{SternPRL80,DasSarma80s,GoldPRL85,GoldPRB86}, Si/SiGe
\cite{GoldPRB87} and III-V \cite{GoldPRB88+90} structures in the
80'th.  In these early works, the resulting $\rho(T)$-dependence
was assumed to be relative small.
The temperature dependence of the resistivity is dominated by the
behavior of the screening function $S(q)$. For $T = 0$, this
function has a distinct kink at $q = 2k_F$ with $k_F$ the Fermi
wave vector. For higher $T$, this kink is smeared out
\cite{SternPRL80} and changes the scattering efficiency especially
for $q = |\vec{k} - \vec{k'}| \approx 2k_F$. Thus for large angle
scattering, the temperature dependence of the screening function
strongly influences the scattering efficiency.

\begin{figure}
\begin{center}
\resizebox{0.85\linewidth}{!}{\includegraphics{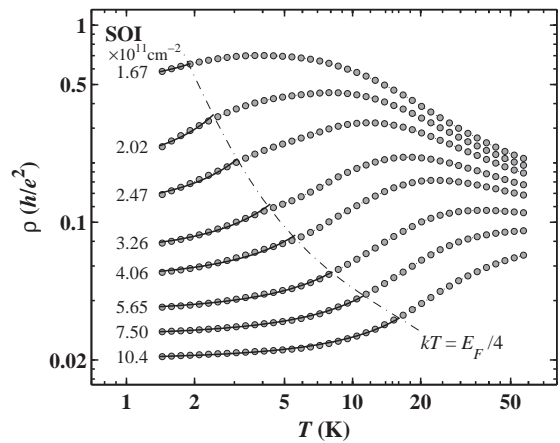}}
\begin{minipage}{8.5cm}
\vspace{0.2cm} \caption{Fit of the temperature dependent
resistivity with $T$ and $T^{3/2}$ screening terms for $k_BT <
E_F/4$. \label{fig:rhoTfitScreening} }
\end{minipage}
\end{center}
\end{figure}

Gold and Dolgopolov give analytic results for the screening
\cite{GoldPRB86} that consist of linear-in-$T$ and $T^{3/2}$ terms
\begin{equation}
  \delta\sigma(T) \approx - C(\alpha,n) \frac{k_BT}{E_f} -
  D(\alpha,n) \left[ \frac{k_BT}{E_f} \right]^{3/2} ,
  \label{eq:dsigmaGold}
\end{equation}
with $C(\alpha,n) = 2C(\alpha)C(n)$, $D(\alpha,n) = 2.45C(\alpha)
\left[ C(n) \right]^2$ and $C(\alpha)$ a constant depending
whether the scattering is dominated by impurities ($\alpha = -1$)
or by surface-roughness ($\alpha = 0$).
We have fitted the $\rho(T)$ behavior with
the Eq.~\ref{eq:dsigmaGold}, assuming $\alpha=-1$ and $C(\alpha)=2\ln2$.
The prefactor $C(n)$ was used as a fitting parameter.
As the analytic results are only valid for $k_BT <
E_F/4$, the fit was limited to that range.  Figure
\ref{fig:rhoTfitScreening} shows that the fit gives a very good
agreement with the observed $T$ dependence.
\vspace{-0.3cm}

\begin{figure}
\begin{center}
\resizebox{0.85\linewidth}{!}{\includegraphics{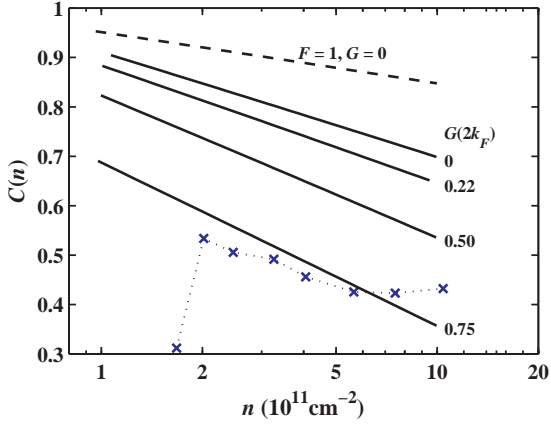}}
\begin{minipage}{8.5cm}
\vspace{0.2cm} \caption{Density dependence of the prefactor $C(n)$
in the screening fit. \label{fig:Cn} }
\end{minipage}
\end{center}
\end{figure}

The fitted prefactor $C(n)$ increases from $0.4$ at $n =
10\times10^{11}$\,cm$^{-2}$ to $0.55$ at
$2\times10^{11}$\,cm$^{-2}$.  For lower densities, the fit does
not give reasonable results as the slope of $\rho(T)$ changes sign
at the critical density $n_c$ which can not be described by the
screening approximation alone. In Fig.\ \ref{fig:Cn}, the range of
the increase in $C(n)$ towards lower densities is compared with
the calculated behavior according to Gold and Dolgopolov
\cite{GoldPRB86}.  As can be seen, there is good agreement with
the curve of the local field correction parameter of $G(2k_F) =
0.75$. But if one inserts $2k_F$ for $q$ in $G(q) =
(1/2g_\nu)q/(q^2 + k_F^2)^{1/2}$ as given for Hubbard's
approximation in Ref.\ \cite{GoldPRB86}, one obtains $G(2k_F) =
0.224$ (with $g_\nu$ the valley degeneracy).  Thus the above
screening description is able to describe the temperature
dependence but has to use a larger value for the local field
correction parameter $G(2k_F)$ as follows from the Hubbard
approximation. This discrepancy possibly results from the fact
that in local field correction parameter only exchange effects are
taken into account and no correlation effects. But the latter
might become important for large electron-electron interaction
parameters $r_s$.  The parameter $r_s$ describes the ratio of
Coulomb to kinetic energy and can be written as $r_s =
g_\nu/a_B\sqrt{\pi n}$, with $a_B$ the Bohr radius.
Recently, Das Sarma and Hwang calculated~\cite{DasSarma99PRL} that
due to temperature dependent screening, changes in $\rho(T)$ by
even up to one order in magnitude can be explained by numerical
calculations.
\vspace{-0.3cm}

\begin{figure} \begin{center}
\resizebox{0.85\linewidth}{!}{\includegraphics{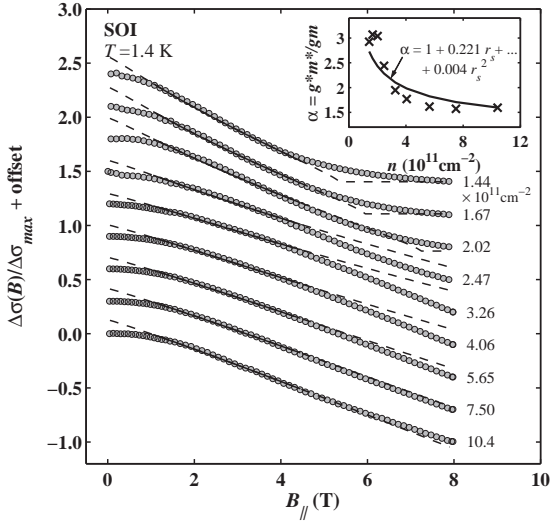}}
\begin{minipage}{8.5cm}
\vspace{0.2cm} \caption{Fit of the SOI magnetoresistivity
$\rho(B_\parallel)$ with the screening behavior according to
Dolgopolov and Gold. Densities are the same as in Fig.
\ref{fig:rhoBpar}. The inset shows the density dependence of the
ratio $\alpha = g^*m^*/gm$
according to the performed fit (x-symbols) and to the literature
(solid line). \label{fig:rhoBparfit} }
\end{minipage} \end{center} \end{figure}

Another puzzling phenomenon in high-mobility Si-MOS structures,
but also in the SOI samples investigated in this work and in other
material systems, is the strong positive magnetoresistance in
parallel magnetic field $B_\parallel$. In Si-MOS structures a more
than ten-fold and in SOI a nearly four-fold increase in
$\rho(B_\parallel)$ was observed.  Dolgopolov and Gold calculated
\cite{Dolgo2000JETPL} that an increase of up to a factor 4 in
$\rho(B_\parallel)$ can take place as the electron system becomes
completely spin polarized.  This strong increase is caused by
changes in the density of states at $E_F$, in the Fermi wave
vector $k_F$ and in the screening wave number $q_s$
\cite{Dolgo2000JETPL}.  We have fitted our data on SOI according
to this theory, the results are shown in Fig.\
\ref{fig:rhoBparfit}.
 \vspace{-0.3cm}

As can be seen, at high electron densities (i.e.\ low
resistivity) the fit in Fig.\ \ref{fig:rhoBparfit} is in good
agreement with the data, whereas at lower densities (i.e.\ high
resistivity and strong changes in $\rho(B_\parallel)$), the
screening behavior leads to curves which are rather linear and do
not show the curvature of the measured values at low $B_\parallel$
fields. For low density, at high $B_\parallel$ fields, the model
calculation shows an abrupt saturation of the resistivity which
corresponds to the complete spin polarization of the electron
system. Both discrepancies between calculation and measurement may
be caused by the fact that the model \cite{Dolgo2000JETPL} was
formulated for $T = 0$\,K. In that case, the Fermi-Dirac
distribution is a step function and leads to the kink at the
$B_\parallel$ saturation field. The approximately linear behavior
of $\rho(B_\parallel)$ for small $B_\parallel$ is probably caused
by the distinct kink \cite{SternPRL80} in the screening function
at zero $T$. For $B = 0$ only one kink exists at $q = 2k_F$ which
splits into two kinks for $B > 0$.\cite{Dolgo2000JETPL} At $T >
0$, the kinks in the screening function are rounded and thus a
small splitting of the two screening functions (by $B_\parallel$)
for the spin up and spin down electron systems will not be so
effective.

In order to fit the screening model to the data, an effective
$g^*$ factor was used as the only fitting parameter instead of a
constant value for $g$. The resulting ratio $\alpha = g^*m^*/gm$
is shown by cross symbols in the inset of Fig.\
\ref{fig:rhoBparfit} in comparison to the ratio $\alpha$ from the
analysis of Shubnikov\,-\,de Haas measurements on Si-MOS samples
\cite{PudalovPRL02b} (solid line).  Over all, there is a good
agreement between $\alpha$ from the screening fit and the Si-MOS
values. Just at the lowest $n$, the deviation is non-monotonic,
but this may be caused by the large deviations in the calculated
and measured $\rho(B_\parallel)$ values as the calculation is
performed for $T = 0$. On the other hand, the above screening
model \cite{Dolgo2000JETPL} is also limited by the Hubbard
approach, which does not take into account electronic correlation
effects. A further work by Gold indicates \cite{Gold2000JETPL}
that even larger ratios of $\rho(B_\parallel \geq B_c)/\rho(B=0)$
than four can be explained beyond the random phase approximation
by taking into account exchange/correlation (i.e.\ many body
effects) and multiple scattering effects .  In this framework
possibly the data for high-mobility Si-MOS samples can be
explained as well.

\subsection{Ballistic interaction corrections}

The fifth and last suggestion (v) on the origin of the metallic
state in 2D concerns coherent interaction corrections (CIC) to the
conductivity in the ballistic regime.
The theory for it was recently introduced by Zala, Narozhny, and
Aleiner (ZNA) \cite{Zala01a}.
Previously, Altshuler and Aronov have calculated quantum
corrections to conductivity (which include especially the
interaction corrections) in the diffusive regime (i.e.\ at low
temperatures, where \ $k_B T \ll \hbar/\tau$) and found a typical
$\ln(T)$ behavior \cite{Altsh85}. For the intermediate $T$ range
(i.e.\ the ballistic regime, where $k_B T > \hbar/\tau$), ZNA
found a linear in $T$ contribution. \cite{Zala01a} In addition,
they calculated the coherent interaction  corrections for the
transition regime, where the behavior changes from
diffusive to ballistic.
In the ballistic regime, the dominant processes can be understood
as coherent backscattering of electrons at the Friedel
oscillations of the other electrons around some perturbation
centers.\cite{Zala01a} In contrast, the coherent scattering in the
diffusive regime occurs at two or more centers into a wider range
of angles. In the approach of ZNA, where the interaction
corrections are calculated in terms of the Fermi liquid parameter
$F_0^\sigma$, correlation (Hartree) and exchange (Fock) terms are
included. As the Fock terms have opposite sign to the Hartree
terms and the relative strength of the two contributions varies
with temperature and electron density, the resulting interaction
corrections may be either positive or negative. The relation of
the ballistic interaction correction to the former screening
approach will be discussed later.

We have fitted the temperature dependence of the SOI-resistivity
by the theory of ZNA, but adopting the situation for two-fold
valley degeneracy in n-type Si inversion layers. As was shown by
Punnoose and Finkelstein \cite{Punnoose01}, the valley degeneracy
leads to 15 spin triplet channels instead of 3 for the one-valley
situation. In both situations there is just on charge channel
participating.
This modification of the ZNA theory was also taken into account in
the evaluation of the resistivity in Si-MOS by Vitkalov et
al.~\cite{VitkalovCM02} and Pudalov et al.~\cite{PudalovCM02}.

\begin{figure} \begin{center}
\resizebox{0.85\linewidth}{!}{\includegraphics{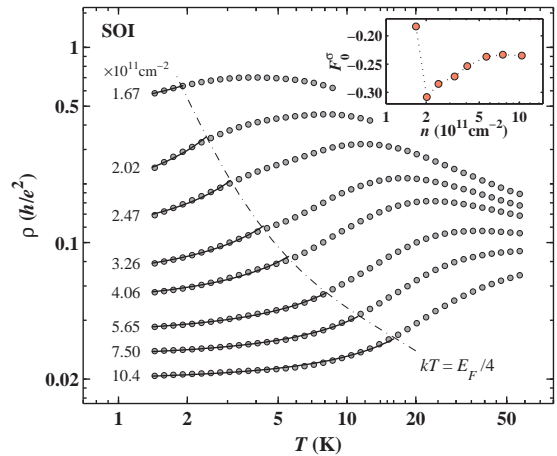}}
\begin{minipage}{8.5cm}
\vspace{0.2cm} \caption{Fit of the temperature dependent
resistivity by the electron-electron interaction corrections in
the ballistic regime according to Zala et al.
\label{fig:rhoTfitZNA} }
\end{minipage} \end{center} \end{figure}

In order to fit the experimental data $\rho(T)$, the functions
$f(T\tau)$ and $t(T\tau;F_0^\sigma)$ for the charge and triplet
channels, respectively, were calculated according to Ref.
\cite{Zala01a} and tabulated. The Fermi liquid parameter
$F_0^\sigma$ and the conductivity $\sigma(T = 0)$ are used as
fitting parameters in a true multi-parameter least-square-fit
procedure. As in the screening case (iv), the fit is performed for
$k_BT \leq E_F/4$. It can be seen from Fig.~\ref{fig:rhoTfitZNA}
that the fitted curve (solid line) describes the experimental
values (circle symbols) quite well. The inset of
Fig.~\ref{fig:rhoTfitZNA} shows the $F_0^\sigma$-values as
obtained from the fitting procedure. In the density range of 2 to
$10.4 \times 10^{11}$\,cm$^{-2}$ the $F_0^\sigma$ values lie
between $-0.23$ and $-0.31$. For the lowest density of $1.67
\times 10^{11}$\,cm$^{-2}$, we find a significantly higher value
of $F_0^\sigma = -0.18$. As can be seen from
Fig.~\ref{fig:rhoTfitZNA}, the fit was performed only in a short
temperature interval and the fit is not so significant there. On
the other hand, one sees also that slop of the $T$-dependence is
indeed drastically changing and one eventually comes into a
different regime of the sample behavior. For even lower electron
densities the fit did not give any reasonable results as the
sample finally becomes insulting below $n \approx 1.5 \times
10^{11}$\,cm$^{-2}$.

In our fitting procedure, we use $\sigma(0)$ as a fitting
parameter. In the work ZNA~\cite{Zala01a} and of Vitkalov et
al.~\cite{VitkalovCM02}, it is suggested to perform a linear
extrapolation of the resistivity $\rho(T)$ towards zero $T$. But
this has the disadvantage that the other contributions (e.g.\
logarithmic ones) may lead to an offset and the value obtained be
the extrapolation towards zero is not the optimal one.  As we have
included the diffusive (i.e.\ logarithmic) terms in our fit, we
find a discrepancy between the fitted value $\sigma(0)$ and the
one from the linear extrapolation. With our procedure, we get a
better consistency for the fit.

 \vspace{-0.3cm}

If we perform our fit on $\rho(T)$ under the assumption of no
valley degeneracy, then the resulting $F_0^\sigma$ values have to
be further negative and are typically in the range of $-0.45$ to
$-0.5$. This comes from the fact that the single charge channel is
always localizing whereas the triplet channels contribute the
delocalizing parts to the conductivity and the strength of the
triplet channels increases towards negative values of
$F_0^\sigma$. If there is just one valley present, the
$F_0^\sigma$ parameter has to be larger in order to lead to the
experimentally observed delocalizing behavior towards low $T$
whereas in the case of two valleys with 15 triplet channels, the
$F_0^\sigma$ value can be lower and still leads to the
delocalizing behavior. For one valley the border between
localizing/delocalizing behavior is given by $F_0^\sigma \approx
-0.4$, for two valleys it is about $-0.2$.  From the quality of
the fit, it has to mentioned that the screening fit of (ii) is
nearly identical, the differences are very small. If one
calculates the prefactor of the linear term in the ZNA theory and
compares it with $C(n)$ from the screening fit, one gets very
similar values.
This means that that in our case the Fock contribution seems to be
not very important.

The interaction corrections at intermediate temperatures were also
calculated for an in-plane magnetic field $B_\parallel$ by
ZNA.\cite{Zala01b} Due to the
bare Zeeman splitting $E_z = g\mu_B B$,
the triplet states with different total spin component $L_z$ are
split and the interference effects (i.e.\ the triplet channels)
are partly suppressed. \cite{Zala01b}
Note that one must not take the renormalized Zeeman splitting
$E_z^* = g^*\mu_B B$, as homogeneous collective modes are not
renormalized by electron-electron interaction.\cite{Zala01b}
We assumed again a two-fold valley degeneracy.  In addition to the
effective Zeeman splitting, a possible valley splitting also
contributes to the conductivity corrections. The expression for
the different corrections is given by \cite{VitkalovCM02}
\begin{equation} \begin{aligned}
  \delta\sigma_{ee} =\ & \delta\sigma_C + 15\delta\sigma_T +
  2\delta\sigma(E_z) + 2\delta\sigma(\Delta_V) \\
  & + \delta\sigma(E_z  + \Delta_V) + \delta\sigma(E_Z -
  \Delta_V),
  \label{eq:dsEzDv}
\end{aligned} \end{equation}
where the terms correspond in the order of
appearance to the charge and triplet channels, to Zeeman and
valley splitting and to combinations of the latter two.

The behavior of the triplet states in parallel field and/or valley
splitting is described by the functions $K_b(x,F_0^\sigma)$ and
$K_d(x/\pi,F_0^\sigma)$ for the ballistic and diffusive case,
respectively. \cite{Zala01b} Here $x = E_x/2T$ and $E_x$ is any of
the energies given as an argument in $\delta\sigma$ of
Eq.~\ref{eq:dsEzDv}.
We have again calculated and tabulated these functions in order to
perform a true multi-parameter least-square fit procedure. First
we describe a fit for $\Delta_V = 0$; the case of $\Delta_V > 0$
is discussed later. For the case of $\Delta_V = 0$, the
conductivity corrections reduce to
\begin{equation}
  \delta\sigma_{ee} = \delta\sigma_C + 15\delta\sigma_T +
  4\delta\sigma(E_z).
  \label{eq:dsEz}
\end{equation}
Each $\delta\sigma(E_z)$ is able to freeze two triplet channels
$2\delta\sigma_T$ (see Eq.~11 in ZNA~\cite{Zala01b}), so that in
the high magnetic field limit $E_z \gg k_BT$ only 7 triplet
channels remain.

\begin{figure} \begin{center}
\resizebox{0.8\linewidth}{!}{\includegraphics{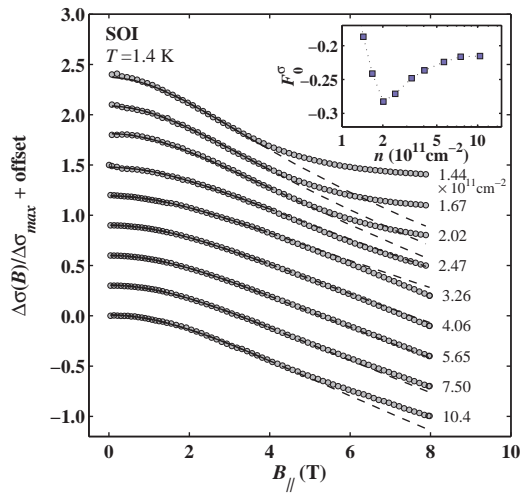}}
\begin{minipage}{8.5cm}
\vspace{0.2cm} \caption{Fit of the SOI magnetoresistivity
$\rho(B_\parallel)$ with electron-electron interaction corrections
in the ballistic regime according to Zala et al. Densities are the
same as in Fig. \ref{fig:rhoBpar}. The inset shows the Fermi
liquid parameter $F_0^\sigma$ (circle symbols) in comparison to
the values calculated from the effective g-factor in Si-MOS (solid
line). \label{fig:rhoBparfitZNA} }
\end{minipage} \end{center} \end{figure}

Figure \ref{fig:rhoBparfitZNA} presents the result of the fit with
$F_0^\sigma$ and $\rho(0)$ the fitting parameters and $\Delta_V =
0$. As can be seen from the figure there is a very good agreement
for high densities over the entire $B_\parallel$-field rang and
also for the lower densities for the low field part. At high
parallel magnetic field and low density there is a discrepancy
between calculations and experiment. This can be understood, as a
limit of applicability of Eq. \ref{eq:dsEzDv} is that $E_z < E_F$.
For the lower electron densities in our experiment, we reach the
total spin polarization regime and the latter relation is not
fulfilled any more. At first we fitted the whole
$B_\parallel$-range, but after recognizing the deviations at high
fields, we limit the fit range to $B_\parallel < 4$\,T as is shown
by the full lines in Fig.~\ref{fig:rhoBparfitZNA}. The
extrapolated $\rho(B_\parallel)$ behavior is indicated by the
dashed part of the lines.

The inset of Fig.~\ref{fig:rhoBparfitZNA} depicts the obtained
values for the fitting parameter $F_0^\sigma$, which are in the
range of $-0.21$ to $-0.28$ for the higher electron densities. For
the lower densities, we find a sudden upturn of $F_0^\sigma(n)$.
It occurs for the same densities, for which there is a large and
systematic deviation at high magnetic fields. Possibly, we reach a
limit of the applicability of the theory.

We have also performed a least square fit with the additional
fitting parameter $\Delta_V$ for the valley splitting. We get
values between $\approx 0$ and 2\,K for $\Delta_V$. These values
do not depend systematically on $n$, but are rather scattered
around in the before mentioned range -- the significance of the
parameter $\Delta_V$ is very low. If we set $\Delta_V = 0$, as
used for the fit above, the deviation of the fitted curve from the
data points is nearly the same and the fitted values of
$F_0^\sigma$ and $\rho(0)$ change only marginally. This means that
we do not get a reasonable estimate for $\Delta_V$ from the fit of
our magnetoresistivity data for in-plane magnetic field.
\\

 \vspace{-0.3cm}

\begin{figure} \begin{center}
\resizebox{0.8\linewidth}{!}{\includegraphics{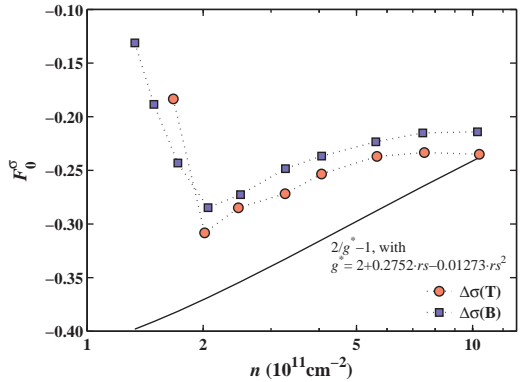}}
\begin{minipage}{8.5cm}
\vspace{0.2cm} \caption{Comparison of the Fermi liquid parameter
$F_0^\sigma$ as obtained from the fit of the temperature
dependence (circles) and the in-plane magnetic field dependence
(squares) of the SOI resistivity.
\label{fig:F_0sig_comparison} }
\end{minipage} \end{center} \end{figure}

Figure \ref{fig:F_0sig_comparison} shows a comparison of the
$F_0^\sigma$ values as obtained from the fits of the $T$- and the
$B_\parallel$-dependence. In the electron density range of 2 to $6
\times 10^{11}$\,cm$^{-2}$, the values (marked by circles and
squares) are close together. This means that two independent
measurements and fits lead to very similar Fermi liquid parameters
$F_0^\sigma$. This observation supports the validity of the
ballistic interaction theory.
For $n < 2 \times 10^{11}$\,cm$^{-2}$, the $F_0^\sigma$ values suddenly increase
strongly. It seems that in this regime the description of the
experiment by the interaction corrections in the ballistic regime
breaks down. This is in good agreement with the observation in
Si-MOS, that for $n < 2 \times 10^{11}$\,cm$^{-2}$, the strong
decrease in $\rho(T)$ towards lower $T$ is in the diffusive
regime, so that the ballistic regime is active only for higher
electron densities.\cite{Bru01PRL}
For comparison, $F_0^\sigma$ was also calculated and plotted for
high-mobility Si-MOS structures (solid line) as obtained
\cite{PudalovPRL02b} from the effective $g^*$-factor by the
relation $g^* = 2/( F_0^\sigma + 1)$. The $F_0^\sigma$ values
obtained by the two different methods show the same trend in that
they increase towards negative values for decreasing electron
density, although the value appears to be more negative for
Si-MOS. This discrepancy is not clear at the moment.

If ones compares the fit of our SOI magnetoresistance data by the
theory on interaction corrections at intermediate
temperatures (ZNA) \cite{Zala01b} with the theory on temperature
dependent screening  (Dolgopolov and Gold) \cite{Dolgo2000JETPL},
it can be seen that the former theory gives a better description
of the experimental data at low magnetic fields. In the first
case, the nearly parabolic dependence of $\rho(B_\parallel)$
around $B = 0$ is described well, whereas in the second case the
fit gives a nearly linear $B_\parallel$-dependence. But as
discussed earlier in this work, this is not astonishing, as the
theory on screening in parallel fields is worked out so far for $T
= 0$, where the screening function has the well defined kink at $q
= 2k_F$.

 \vspace{-0.3cm}

\subsection{On a possibly new ground state}

The first suggestion (i) on a new quantum mechanical ground state
in the metallic regime should be discussed once more in the light
of screening and coherent electron transport in the ballistic
regime. In a recent work \cite{Bru01PRL}, we concluded that the
metallic state in Si-MOS samples is not caused by quantum coherent
effects. We discussed screening as a prominent candidate for the
metallic effect. At this time screening was not considered as a
quantum coherent effect, primarily. In the new sight of ZNA,
screening can be described by coherent back scattering of
electrons at the perturbation centers and it's Friedel
oscillations. It seems like suddenly the effect has changed it's
habits.

If we now try to describe the strong resistivity decrease in a
combined picture of screening and coherent ballistic contributions
(let's denote it coherent screening regime), one should say that
the correlation (Fock) contribution may give an important
contribution and was missed in the former screening descriptions
\cite{SternPRL80,DasSarma80s,GoldPRB86}. Only in some later work,
the correlation contributions were included in the local field
contribution \cite{Gold2000JETPL,Gold01JPhysC}. On the other hand,
the effect still reflects the behavior of the screening function
with it's sharp kink at $T = 0$ which is rounded for $T > 0$. In
any case, the coherent screening effects leads to a linear $T$
behavior and no new or extraordinary effects will emerge from
itself.

The important realization is that the coherent conductivity
corrections in the diffusive regime can be understood as a natural
continuation of the coherent screening. It becomes important as
soon as the coherent scattering extends over more than one
scattering center at a time (which is known since many years). The
diffusive regime describes the weak localization effect and the
coherent disorder-induced electron-electron interaction
corrections. \cite{Altsh85} Only if the latter contributions lead
to strong and dominant conductivity corrections, one should speak
of a new quantum mechanical ground state. Such a scenario was
described by Punnoose and Finkelstein \cite{Punnoose01} in the
case of two degenerate valleys where the delocalizing effect of
the 15 triplet channels can be significantly enhanced due to
renormalization effects. But this is not the case for higher
electron densities of $n > 2 \times 10^{11}$\,cm$^{-2}$ as was
shown in the recent work on the borders for quantum effects in the
diffusive regime. \cite{Bru01PRL} Only for lower electron
densities, such a renormalization effect could significantly
increase the delocalizing contributions and lead to a new ground
state.

\section{Summary}
In conclusion, we have shown that silicon-on-insulator structures
show a very similar 2D-metallic state as Si-MOS structures with
comparable peak mobility.  We discussed and compared several
models on the physical origin of the metallic state. The theory
on coherent interaction corrections at intermediate temperatures
is able to describe the observed temperature and in-plane magnetic
field behavior over a wide density range. Under the assumption of
a two-fold valley degeneracy, the Fermi-liquid parameter
$F_0^\sigma$ is between $-0.21$ and $-0.31$. The derivated
Fermi-liquid parameters from the temperature and the in-plane
magnetic field dependence are in very good agreement to each
other.  This observation points to the validity of the ballistic
interaction theory.  We do not get a significant influence from
the valley splitting parameter $\Delta_V$. The earlier screening
approach
gives also a similarly good description for intermediate
temperatures. This indicates that the correlation (or Fock) terms
may not be very significant in the investigated regime.  The
charged trap state model, as given in the literature,
is not able to describe the resistivity versus density behavior
further away from the critical density of the metal-to-insulator
transition. Also the two-band transport model seems to be not
applicable for electrons in Si-layers.  It is not clear yet
whether there exists a true metallic state for low temperatures at
low electron densities or not.

\section*{Acknowledgments}
We thank B.\,L.~Altshuler, I.~Aleiner, V.\,T.~Dolgopolov,
K.~Ensslin, A.~Finkelstein, A.~Gold, B.~Narozhny, D.\,L.~Maslov,
and V.\,M.~Pudalov for stimulating discussions.
The work was supported by the Austrian Science Fund (FWF) on the
``Metallic State in 2D Systems''.

\appendix
\section{Trap-model calculations} \label{sec:AppendixTrapModel}

Analytic results for the metal-insulator transition in gated
semiconductors according to trap states were calculated recently
by Altshuler and Maslov (AM) \cite{Altsh99PRL}.
In this work, AM assume that a hole trap exists in the oxide
layer, which can easily be charged and discharged. The trap state
is homogeneously distributed in space but has a distinct
energetical position.
Some of the equations used by AM were not derived in that work
according to the length limits of the letter publication and in
addition several approximations were used there.
We find it therefore useful to discuss the calculations and our
numerical results in more detail here.

\begin{figure} \begin{center}
\resizebox{0.8\linewidth}{!}{\includegraphics{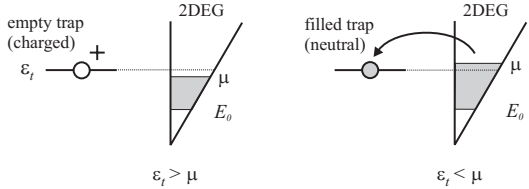}}
\begin{minipage}{8.5cm}
\vspace{0.2cm} \caption{Schematic representation of a hole trap
center in relation to the 2D electron gas. For $\mu <
\varepsilon_t$ the trap state is positively charged whereas for
$\mu > \varepsilon_t$ the trap are neutral and do not act as
scattering centers.
\label{fig:trap_scheme} }
\end{minipage} \end{center} \end{figure}

Figure \ref{fig:trap_scheme} shows a simple scheme of the hole
trap states. If the chemical potential $\mu$ (we use here now the
same notation as by AM~\cite{Altsh99PRL}) is below the trap state
energy $\varepsilon_t$ (Fig. \ref{fig:trap_scheme}a), a hole is
trapped, it is positively charged and acts as an efficient
scattering center. For the case that $\mu$ is above
$\varepsilon_t$, the center is neutral and is inefficient for
scattering.

 \vspace{-0.3cm}

\begin{figure} \begin{center}
\resizebox{0.85\linewidth}{!}{\includegraphics{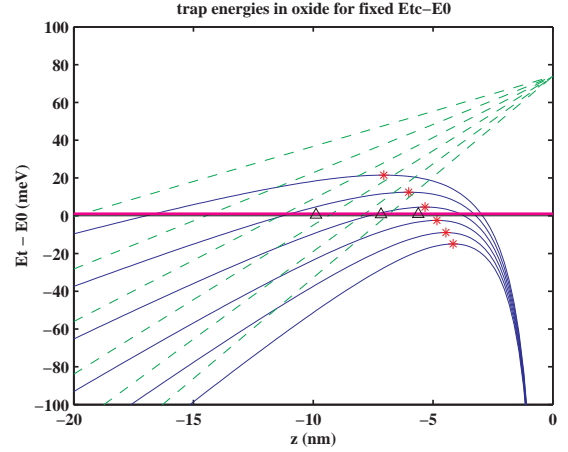}}
\begin{minipage}{8.5cm}
\vspace{0.2cm} \caption{Full lines show trap energy vs.\ distance
$z$ from the 2D electron gas layer plotted for different gate
voltages $V_g$.  The dashed lines indicate the potential energy of the
trap without the mirror charge term; they merge on the right hand
side at the trap energy $\varepsilon_{t0}$. The thick horizontal
line indicates the position of the Fermi energy $E_F$; it actually
consists of several thin lines for different values of $E_F(V_g)$,
but as the Fermi energies are so small in comparison to the
considered trap energies, they are not resolved. The stars depict
the maxima of the energy trap curves, the triangles show the mean
$z$-value of the positive region, i.e.\ where $\varepsilon_t >
\varepsilon_0 $ with $\varepsilon_0$ the lowest 2D energy level in
the Si inversion layer.
\label{fig:trap_energy}}
\end{minipage} \end{center} \end{figure}

For the scattering centers, a dipole approximation is used by AM.
It is assumed that a charge in the oxide layer will introduce an
opposite mirror charge in the 2D layer due to screening on
in-plane electric fields. For the energetical position of the trap
state, a term $e^2/(8\pi\epsilon_0\epsilon_{\rm ox}z)$ has to be
taken into account, with $\epsilon_0 = 8.854 \times
10^{12}$\,C$^2/$Nm$^2$ the vacuum permeability, $\epsilon_{\rm
ox}$ the relative dielectric constant of the oxide and $z$ the
distance of the charge from the 2D electron layer; the
calculations are performed in SI units here. The energetical
position of the charged trap states also depends on the applied
gate voltage via $eV_gz/d$, with $d$ the thickness of the oxide
layer. Figure \ref{fig:trap_energy} depicts the dependence of the
total energy of the charged trap states in the dipole approximation
as a function of the distance $z$ for different gate voltages
$V_g$. Only in the spacial region where $\varepsilon_t(z) > \mu$,
the centers are positively charged.

The expression for the resistivity $\rho$ follows in the
Drude-Boltzmann approximation by calculating the effective
transport scattering time
\begin{equation}
  \left\langle\tau\right\rangle =
  \frac{\int{d \varepsilon\, \tau(\varepsilon) \varepsilon \, \partial f/\partial \varepsilon}}
  {\int{d \varepsilon\, \varepsilon \, \partial f/\partial \varepsilon}}
  \label{eq:tau}
\end{equation}
for the conductivity $\sigma = ne^2 \left\langle\tau\right\rangle
/m^* $. The weighting of $\left\langle\tau\right\rangle$ with the
kinetic electron energy $\varepsilon$ in Eq.\ \ref{eq:tau}
fundamentally follows from the Drude-Boltzmann approximation
\cite{Smith}, as the Fermi velocity $v_F$ and the shift of the
Fermi surface in k-space are both proportional to
$\sqrt{\varepsilon}$, which enter into the expression for the
current $j_x = -e \int{d \vec{k}\, n(\vec{k})f(\varepsilon) v_x
}$. The integral in the denominator of Eq.\ \ref{eq:tau} is in 2D
just equal to the Fermi energy $E_F$ -- also for elevated
temperatures.

 \vspace{-0.3cm}

The transport scattering time $\tau(\varepsilon)$ has to be
calculated from the scattering rate
\begin{equation}
  1/\tau(\varepsilon) = \int_{0}^{d}{ dz\, N_{t3}^+(z)\, v(\varepsilon)\,
  \sigma_{\rm scm}(\varepsilon,z)}
  \label{eq:scat_rate}
\end{equation}
with $N_{t3}^+(z)$ the three dimensional density of charged traps,
$v(\varepsilon)$ the electron velocity and $\sigma_{\rm
scm}(\varepsilon,z)$ the \underline{s}cattering \underline{c}ross
section for \underline{m}omentum relaxation. According to AM, the
scattering cross section for a dipol is given by $\sigma_{\rm
scm}(\varepsilon,z) =
2.74(e^2z^2/8\pi\epsilon_0\epsilon^*\varepsilon)^{1/3} $. The
effective relative permitivity $\epsilon^* = (\epsilon_{\rm Si} +
\epsilon_{\rm ox})/2 \approx 7.9$ is the mean value of the Si and
the SiO$_2$ layer. The density $N_{t3}^+(z) = N_{t3} P_+(z)$ in
the AM notation with $N_{t3}$ a constant density of traps per
volume and $P_+(z) = [0.5
\exp((\mu-\varepsilon_e(z)-\varepsilon_t)/k_BT)+1]^{-1}$, the
thermodynamic probability for a trap state to be
charged\cite{Altsh99PRL}.

By inserting the above expressions into Eq.\ \ref{eq:scat_rate}
one gets
\begin{equation}
  1/\tau(\varepsilon) = c' N^+_{\rm eff} z_m^{2/3} \varepsilon^{1/6}
  \label{eq:scat_rate2}
\end{equation}
with the factor $c' = 2.74(e^2/8\pi\epsilon_0\epsilon^*)^{1/3}
\sqrt{2/m^*}$, an effective number of positive trap states per
area $N^+_{\rm eff} = N_{t3} \left\langle \Delta z
\right\rangle_{\rm eff}$, the effective width of positive charge
layer $\left\langle \Delta z \right\rangle_{\rm eff} = \int{{\rm
d}z P_+(z) (z/z_m)^{2/3} }$ and the position $z_m =
\sqrt{ed/8\pi\epsilon_0\epsilon_{ox} V_g}$ of the energetical
maximum of the trap energy \cite{Altsh99PRL}. By inserting Eq.\
\ref{eq:scat_rate2} into Eq.\ \ref{eq:tau}, one gets
\begin{equation}
\left\langle \tau \right\rangle \propto 1/\mu \int_{0}^{\infty}{
d\varepsilon\,
\varepsilon^{5/6} \partial f/\partial \varepsilon} .
\label{eq:tau_eff}
\end{equation}
For the Drude resistivity $\rho = m^*/ne^2 \left\langle \tau
\right\rangle$, an effective energy $\bar{\varepsilon}$ can be
defined so that a relation equivalent to Eq.~\ref{eq:scat_rate2}
can be written for the effective values, i.e.\ $1/\left\langle
\tau \right\rangle = c' N^+_{\rm eff} z_m^{2/3}
\bar{\varepsilon}^{1/6}$. A simple calculation gives
\begin{equation}
\bar{\varepsilon} = \mu^6 \left[\int_{0}^{\infty} d\varepsilon\,
\varepsilon^{5/6}\,
\partial f/ \partial \varepsilon \right]^{-6} .
  \label{eq:Eeff}
\end{equation}
The first derivative of the Fermi-Dirac function $f$ can be
expressed as
\begin{equation} \begin{aligned}
\partial f / \partial \varepsilon =& - f (1-f)\\ =& -(4k_B T
\cosh^2((\varepsilon - \mu)/2k_B T))^{-1}
\label{eq:df_dE}
\end{aligned} \end{equation}
and one obtains the same expression as Eq.\,8 in
Ref.~\cite{Altsh99PRL} With these relations, the resistivity can
exactly be written in terms of the effective energy
$\bar{\varepsilon}$ as
\begin{equation}
  \rho = \frac{m^*}{ne^2} N^+ v(\bar{\varepsilon})
  \sigma_{\rm scm}(\bar{\varepsilon},z_m) \, .
  \label{eq:rho(E)}
\end{equation}

We have calculate the resistivity $\rho(n,T)$ by numerical
integration for the effective thickness $\Delta z$ of the layer
with charged trap states and the effective energy
$\bar{\varepsilon}$. In the work of AM \cite{Altsh99PRL}, a
parabolic approximation was used for the dependence of the trap
energy $\varepsilon_t$ on $z$ and an analytic expansion was given
for the behavior of $\rho(n,T)$. With this, AM were able to show
that in principle a critical behavior in the density dependence of
$\rho(T)$ can appear. As in our work, we want to test whether the
metallic behavior can be explained for a wide density range or
not, we use the numerical calculation method in order to be able
to perform a precise evaluation of the model behavior for large
deviations of
$n$ from the critical density $n_c$.

AM proposed in their model that the dependence of $\rho(n)$ in the
vicinity of $n_c$ comes from the strong change of the number of
positively charged centers, as the Fermi energy is just above or
below the energy maximum $\varepsilon_m$ of the trap states, where
they have a singularity in the density of states. The sensitivity
on temperature comes from the broadening of the thermodynamic
probability distribution $P_+$ with increasing temperature. In
order to test this behavior for the strong temperature dependence
of the resistivity, i.e.\ the most characteristic feature of the
metallic state, we calculated the temperature dependence within
the model at an intermediate temperature of $T = 3$\,K. For
densities $n < 5 \times 10^{11}$\,cm$^{-2}$, $\rho(T)$ does not
saturate yet at its low-$T$ value. The saturation behavior is not
described by the AM-model. The low-$T$ value has to be cause by
other mechanisms like residual impurities which do not change
their charge state in the same manner. For the evaluated value of
$T \approx 3$\,K, the $T$-dependence should behave according to
the AM-model if it is able to describe the metallic state.

\begin{figure} \begin{center}
\resizebox{0.80\linewidth}{!}{\includegraphics{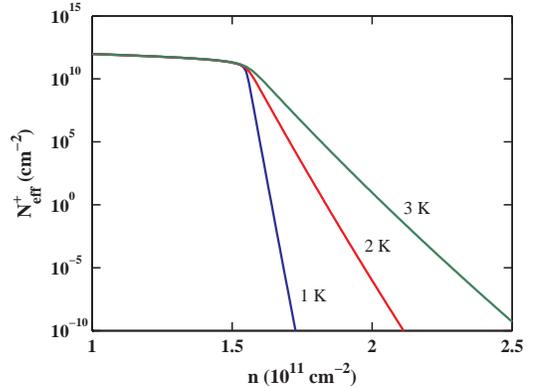}}
\begin{minipage}{8.5cm}
\vspace{0.2cm} \caption{Effective number of charged trap states
per area $N^+_{\rm eff}$ vs.\ electrons density (i.e.\ varying
gate voltage) at temperatures of $T = 1$, 3, and 5\,K.
\label{fig:N+eff(n)} }
\end{minipage} \end{center} \end{figure}

The result of our numerical calculation is shown in Fig.\
\ref{fig:rho_trap_SOI}. It is clearly visible there, that the
calculated $\rho$-dependence changes by many more orders of
magnitude than the measured one. In order to analyze the origin of
the strong change in the calculated $\rho(n)$ behavior, we show in
Fig.\ \ref{fig:N+eff(n)} the dependence of the effective
number of charged trap states per area $N^+_{\rm eff}$
on the electron density (i.e.\ the gate voltage $V_g$). As long as
$\varepsilon_m > \mu$ a part of the trap states is positively
charged even for $T = 0$ and the variation in effective positive
charge layer width $\Delta z$ is small. But for $\varepsilon_m <
\mu$, the trap states are only charged according to the
exponentially decaying part of the thermodynamic probability
function $P_+(z)$ and thus the strong exponential $T$-dependence
occurs.

In the same density range, the other values which enter into the
resistivity calculation (see Eq.\ \ref{eq:rho(E)}) change only by
about a factor of two. It is thus clear that within the proposed
AM model, the strong shift of the energetical position of the trap
states with the applied gate voltages leads to too strong changes
in $\rho(n)$ further away from the critical density $n_c$.

\end{multicols}
\end{document}